\documentclass[conference]{IEEEtran}
\IEEEoverridecommandlockouts
\usepackage{cite}
\usepackage{amsmath,amssymb,amsfonts}
\usepackage{algorithmic}
\usepackage{graphicx}
\usepackage{textcomp}
\usepackage{xcolor}
\usepackage{todonotes}
\usepackage{soul,color}
\renewcommand{\hl}[1]{#1}
\usepackage{makecell}

\def\BibTeX{{\rm B\kern-.05em{\sc i\kern-.025em b}\kern-.08em
    T\kern-.1667em\lower.7ex\hbox{E}\kern-.125emX}}
\begin{document}

\title{Efficient and Secure Discovery for AI Agents: \\
The Case for DNS}

\author{
\IEEEauthorblockN{Ramachandra Rao Seethiraju}
\IEEEauthorblockA{
Verisign\\
rseethiraju@verisign.com
}
\and
\IEEEauthorblockN{Sameer Thakar}
\IEEEauthorblockA{
Verisign\\
sthakar@verisign.com
}
\and
\IEEEauthorblockN{Karthik Shyamsunder}
\IEEEauthorblockA{
Verisign\\
kshyamsu@verisign.com
}
\and
\IEEEauthorblockN{Eric Osterweil}
\IEEEauthorblockA{
Verisign\\
eosterweil@verisign.com
}
}

\maketitle

\begin{abstract}

As Artificial Intelligence (AI) agents enter their next stage of ubiquitous deployment throughout the Internet, their discoverability will become a central challenge.
The information AI agents need to discover one another, how they will locate it, how to facilitate authentication, integrity, and authorization, how to connect across different platforms, and how to scale across organizational boundaries form a set of unanswered challenges that will be central to deployment success.
These are challenges for which one of the Internet's most venerable, solid, and  ubiquitous infrastructures is ideally suited: The Domain Name System (DNS). 
Such a rich, already ubiquitous, and programmatically flexible foundation is an ideal option for discovery of AI agents.
In this work, we propose an illustration and rationale for the basic semantics that discovery for AI agents will require.
We argue that three key evaluation criteria will become paramount: 
navigational completeness (the extent to which the necessary metadata, with elements like trust, is included in a discovery solution),
lookup complexity, 
and transaction performance (e.g., latency, speed, or recency). 
Using 119,757 real-world service endpoints, 62,739 indexed MCP server entries, and multiple agent tooling ecosystems, we empirically evaluate the first of these considerations to illustrate the appropriateness of using DNS for AI agent discovery. 
\hl{We find that the necessary and sufficient discovery metadata can be encoded into a single DNS resource record that fits within one DNS UDP response, enabling single-RTT discovery.}
Our evaluations illustrate a promising path toward enabling AI agent discoverability at the Internet's scale, and thereby accelerating secure, stable, and resilient deployments.
\end{abstract}


\section{Introduction}

As Artificial Intelligence (AI) agents grow rapidly on the Internet, their 
impact is already rattling some Internet foundations.
AI agents are automating actions, discovering data, and discovering each other, sometimes \emph{without humans in the loop.}
However, their initial settings have largely been in constrained ``closed-world'' deployments.
Evolving and deploying them onto the Internet necessarily introduces many basic challenges that AI agent discovery has not faced previously.
This frames one of the most foundational challenges: 
how can AI agents search, hypothesize, and optimize across science, engineering, and infrastructure?
We argue that secure discovery is a broad topic, whether AI agents are discovering data, agents, or agents that operate on data.
Additional challenges exist, but we propose that establishing a general approach for secure AI agent discovery is a necessary first step.

As AI agents begin to grow out of ``closed-world'' deployments, they will face axiomatic challenges.
In these closed ecosystems, named AI agents can homogeneously discover each other through their native platforms.
For example, Claude agents can discover other Claude agents by name, and ChatGPT agents can discover other ChatGPT agents in a similar manner. 
However, these agents do not have an inherent mechanism to discover and interoperate heterogeneously across platforms (e.g., Claude-to-ChatGPT).

AI agents in homogeneous settings discover each other by \emph{name}, and through closed system (i.e., platform) registries of names.
This does not immediately befit the Internet setting.
For any agent to discover any other agent at the Internet's scale, agents will need names on the Internet.
This is the function of the Domain Name System (DNS)~\cite{mockapetris1988development}.
However, beyond just names, AI agents need to be associated with capabilities, be able to authenticate, establish integrity protections, identify service endpoints, and locate security status mechanisms.
Work to address these considerations is underway in the Internet Engineering Task Force (IETF) and other communities.
Notable examples of this work are DNS for AI Discovery (DNS-AID)~\cite{mozleywilliams-dnsop-dnsaid-01}, and the Agent Name Service (ANS)~\cite{narajala-courtney-ansv2-01}, as well as a growing list of others~\cite{dawn-list}.
However, which of these candidate proposals provide the most appealing solutions, how well do they fit AI agents' requirements, and how can they be \emph{evaluated?}

In this work, we propose a baseline set of considerations for using DNS to enable AI agent discovery on the Internet.
%
%
We propose an evaluation framework that identifies a set of basic metadata needed for association and three performance evaluation criteria.
These criteria are navigational completeness (which centers on metadata size), lookup complexity, and transaction performance.
%
%
Our early results of the metadata size criterion show that the DNS, its Security Extensions (DNSSEC)~\cite{rfc4033,rfc4034,rfc4035}, together with DANE-like certificate binding mechanisms (e.g., the DNS-Based Authentication of Named Entities (DANE)~\cite{rfc6698,osterweil2012reducing}), can hit a sweet spot.
%
%
%
\hl{Our evaluation demonstrates that the necessary and sufficient discovery metadata can fit within  DNS' deployment constraints.}
\section{State of Play}\label{sec:bg}

The DNS is a valuable, well developed, and well understood inter-domain namespace and infrastructure.
It is a general distributed database that allows globally unique names to map to arbitrary service identifiers~\cite{osterweil2012behavior}.
In addition to IP addresses, these identifiers include cryptographic keys, policy directives for email, and service bindings (e.g., the {\tt SVCB} record~\cite{rfc9460}).
This set of data types, and the DNS itself, are actively being extended.
Recent work to reconsider DNS' delegation structure (the deleg working group~\cite{deleg-wg}) initially considered {\tt SVCB} records as a foundation but 
settled on a new record type with some common features.
At its core, however, the DNS is designed to operate at extremely high speeds, where query/response transactions are measured in milliseconds.
Coupled with this, DNSSEC adds security assurances that cryptographically bind domain names to their rightful origins and protect their integrity~\cite{osterweil2009deploying}.

These reasons underscore the growing interest in using DNS for AI agent discovery.
For example, DNS-AID constructs AI agents' names as DNS domain names that include the agents' capabilities, and requires DNSSEC. 
Those names are then mapped to DNS service binding records ({\tt SVCB}).
DNS-AID then encodes other necessary metadata and pointers to metadata into those records.
This includes service endpoints.
In addition, DNS-AID uses DANE {\tt TLSA} records to establish origin authentication, integrity, and confidentiality protections for the AI agents' specified endpoints.
In DNS-AID, discovering agents issue multiple queries for metadata, and for TLS authentication information.

Another recent approach, ANS, also makes use of DNS domain names as the foundation for agent identity and trust.
As with DNS-AID, ANS distributes the metadata necessary for agent discovery across multiple DNS records and artifacts, requiring discovering agents to assemble the metadata needed for interaction.
Its design combines DNS-based records with additional metadata artifacts and trust infrastructure. 
In summary, across the growing innovation space of AI agent discovery, we observe varied sets of metadata being encoded, variable numbers of lookups and retrievals required, differing dependencies on external artifacts and services, and potentially diverse ranges of network transaction overheads.
\section{A Framework for Evaluating Cross-Platform AI Agent Discovery}

The breadth and scope of the landscape of AI agent discovery proposals is large and growing.
This clearly frames the need for our work: a framework for evaluating \emph{considerations}.
To construct this framework, we first identify the necessary \hl{and sufficient discovery} metadata \hl{required} to enable AI agent discovery at the Internet scale. \hl{
This is the 
information required to locate, verify, and initiate communication with a 
discovered
agent. 
Our framework 
evaluates discovery performance after an agent has been identified or selected. 
It also excludes protocol-specific runtime metadata exchanged after communication begins. Accordingly, navigational completeness evaluates the necessary and sufficient discovery metadata, separate from runtime interaction metadata.}

To do this, we begin by synthesizing several prominent current approaches, such as~\cite{mozleywilliams-dnsop-dnsaid-01, narajala-courtney-ansv2-01}.
From this synthesis, we extract that, for {\bf \emph{``navigational completeness,''}} AI agent discovery needs names to be \emph{``associated''} with a core set of metadata \hl{that} enables:
identifying and determining where another agent is located (\emph{locatability}), being able to interpret what functionality it provides (\emph{capability awareness}), understanding how to communicate with it (\emph{protocol awareness}), and establishing the authenticity and integrity of both the retrieved metadata and the actual connection with the agent. 
Across the different discovery proposals, these necessary metadata elements are encoded and served differently, which leads to different lookup requirements.

The next aspect of our evaluation framework is the {\bf \emph{``lookup complexity.''}}
This involves the requirements for agents to discover and obtain the sufficient set of metadata needed.
For example, the ANS approach requires a DNS lookup (or ``resolution'') for the domain name component of an agent's ANS name.
Then it requires additional lookups for the agent's DANE ({\tt TLSA}) and several other records.
It further requires a lookup to the specified ANS agent registry.
Alternately, the DNS-AID approach first requires a single lookup for an agent's name, which retrieves a corresponding {\tt SVCB} record.
Then, however, it requires a DANE lookup and, because some of the elements in the {\tt SVCB} refer to external registries, it requires further lookups before a connection can be established to a discovered agent.
Looking these elements up in separate registries, and potentially over separate types of transport protocols, increases the complexity of this step.

Finally, in concert with the lookup complexity, the overall reliability and efficiency of the {\bf \emph{``transaction performance''}} of AI agent discovery forms the third component of our framework.
Our synthesis in this work reveals a tradeoff that different discovery approaches may implicitly make.
Some approaches involve combinations of UDP-centric protocols (like the DNS) and TCP-centric protocols (like HTTPS).
In addition to the complexity of requiring multiple lookups, requiring multiple round-trip times (RTTs)  necessarily also affects the performance and latency. These additive RTTs may vary in performance because some of the above transactions over DNS (e.g., the DANE and {\tt SVCB} queries) may query separate name servers, some may involve slower HTTPS lookups, etc.
Our framework illustrates that these decisions will have quantitative impacts on the overall performance and scalability of solutions.
\hl{Taken together, these three elements form the basis of our evaluation framework.}
\section{Evaluation of DNS as a Case Study}
\label{sec:eval}
\hl{As an initial evaluation of our framework, we focus on \emph{navigational completeness}, which focuses on the size of the necessary and sufficient discovery metadata. This has become a driving issue in ongoing Internet Engineering Task Force (IETF) discussions on DNS-based AI agent discovery.
}

Due to the scarcity of live AI agent \hl{datasets containing both deployed service endpoints and capability metadata, we approximate discovery metadata using complementary real-world datasets.} 
\hl{We derived endpoint sizes from API endpoint URLs in the APIs.guru dataset}~\cite{apisguru-directory,apisguru-index}, \hl{capability identifiers from the MCPZoo}~\cite{wu2025mcpzoolargescaledatasetrunnable}, \hl{and protocol identifiers from a limited set of interaction protocols (e.g., MCP and A2A).}

\begin{figure}[t]
\centering
\includegraphics[width=\linewidth]{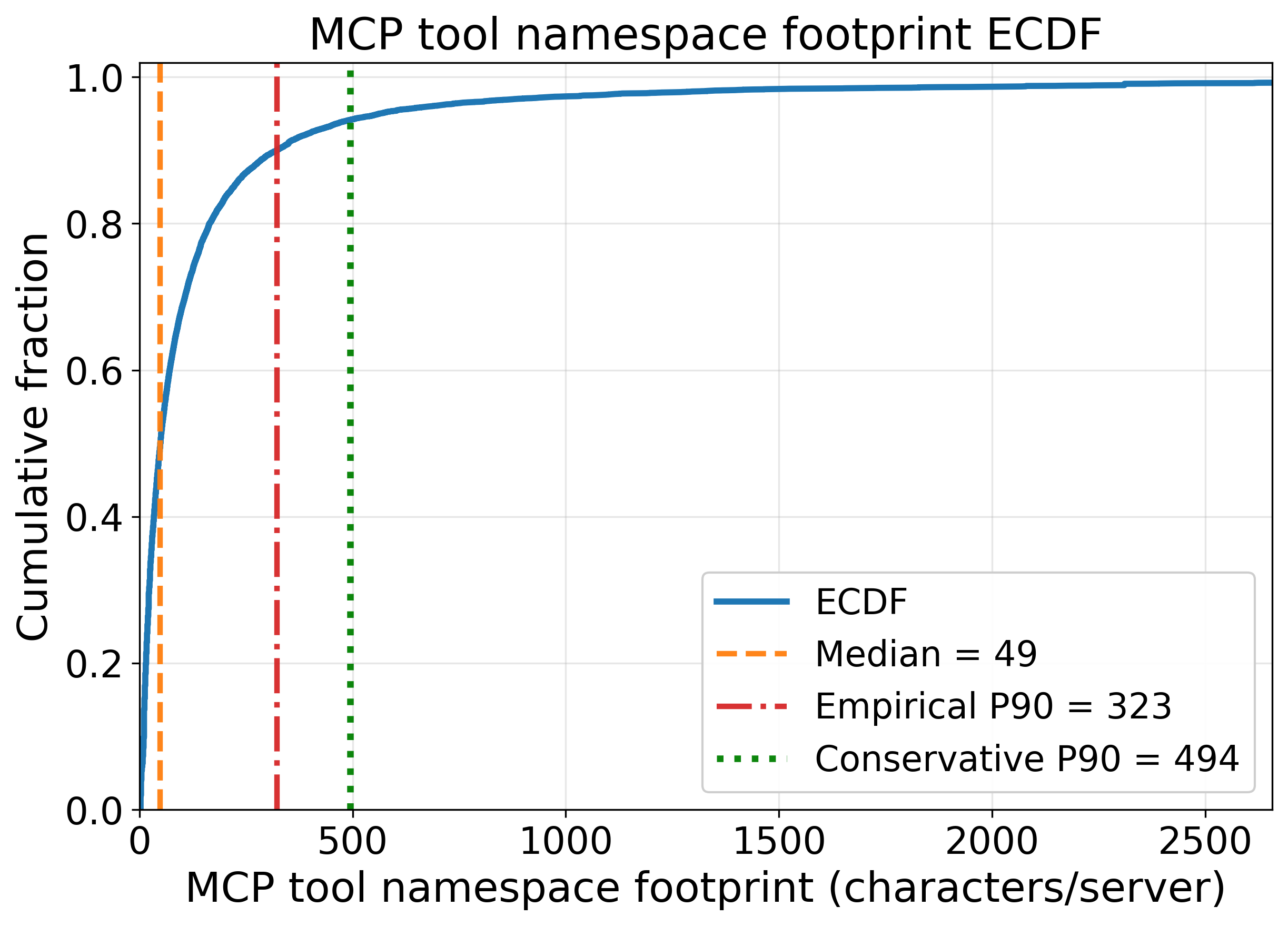}
\caption{Distribution of MCP capability footprint. Empirical P90: 323 characters/server; sizing analysis uses a conservative upper bound of 494.}
\label{fig:capability-ecdf}
\end{figure}

From the APIs.guru dataset we extract 143{,}634 endpoint observations, reduced to 119{,}757 unique endpoints after deduplication. \hl{The endpoint-length distribution has a median of 70 bytes and a 90th percentile of 108 bytes} (Table~\ref{tab:metadata-stats}).

\begin{table}[t]
\caption{\hl{Summary Statistics for Endpoint and Capability Metadata.}}
\label{tab:metadata-stats}
\centering
\setlength{\tabcolsep}{4pt}
\begin{tabular}{lccccc}
\hline
\textbf{Metric} & \textbf{P50} & \textbf{P90} & \textbf{P95} & \textbf{P99} & \textbf{Max} \\
\hline
\makecell[l]{Endpoint length (bytes)}
    & 70 & \textbf{108} & 118 & 143 & 283 \\

\makecell[l]{Tools per MCP server}
    & 3 & \textbf{19} & 35 & 125 & 395 \\

\makecell[l]{Tool name length (characters)}
    & 16 & \textbf{26} & 29 & 37 & 65 \\

\makecell[l]{Capability footprint (characters/server)}
    & 49 & \textbf{323} & 572 & 1918 & 15065 \\
\hline
\end{tabular}
\end{table}

To evaluate capability size requirements, we analyzed 62,739 indexed MCP server entries from MCPZoo~\cite{wu2025mcpzoolargescaledatasetrunnable} and extracted MCP tool declarations directly from source code using pattern-based parsing, without executing the MCP servers at runtime. \hl{This methodology characterizes the discovery metadata published by deployed MCP server implementations rather than their dynamic runtime behavior. Consequently, our analysis reflects the metadata available prior to interaction, consistent with the discovery-focused scope of this evaluation.} In our evaluation, we treat MCP tool identifiers as operational capability identifiers exposed by deployed AI, since they represent externally discoverable and invokable agent functionality. 
\hl{Accordingly, we use MCP tool identifiers as a proxy for coarse capability awareness at  discovery and not to represent the richer protocol-specific information exchanged during runtime.}

We found that 13,607 servers (21.7\% of the corpus) contained at least one detectable MCP tool declaration. Table~\ref{tab:metadata-stats} \hl{summarizes the observed distributions of tools per server, tool-name length, and capability footprint. The empirical P90 capability footprint is 323 characters per MCP server} (Figure~\ref{fig:capability-ecdf}).

\hl{For sizing, we
needed to combine data from our two independent sets: endpoints and MCP servers.
For this, we
used the empirical P90 endpoint length (108 bytes) together with a conservative capability upper bound of 494 characters per server. 
We obtained this upper bound by combining the observed P90 MCP tool count (19) and tool-name length (26). Because these percentiles were measured independently, the resulting value was a conservative upper bound, rather than the empirical P90 capability footprint (323 characters). Combining the endpoint length (108 bytes), capability bound (494 bytes), and protocol identifier (7 bytes) yielded a subtotal of} \textbf{609 bytes} (Table~\ref{tab:size-breakdown}).

\begin{table}[t]
\caption{\hl{Conservative DNS Message-Size Accounting.}}
\label{tab:size-breakdown}
\centering
\begin{tabular}{lr}
\hline
\textbf{Component} & \textbf{Bytes} \\
\hline
Endpoint metadata (P90) & 108 \\
Capability footprint (upper bound from independent  P90s) & 494 \\
Protocol identifier & 7 \\
\hline
\textbf{Discovery metadata subtotal} & \textbf{609} \\
DNSSEC overhead (conservative RSA-2048) & +296 \\
\hline
\textbf{DNS response with DNSSEC} & \textbf{905} \\
Certificate association (SHA-256) & +35 \\
\hline
\textbf{DNS response with DNSSEC and certificate association} & \textbf{940} \\
DNS message header & +12 \\
Representative DNS name encoding (pre-compression) & +60 \\
DNS protocol fields (type, class, TTL, RDLENGTH) & +14 \\
\hline
\textbf{Estimated DNS message size} & \textbf{1026} \\
\hline
\end{tabular}
\end{table}

Applying DNSSEC ensures authenticity and integrity of the domain names. 
To evaluate the impact of this, we include RRSIG overhead based on the record structure defined in RFC 4034~\cite{rfc4034} and signature sizes specified in RFC 6605~\cite{rfc6605}. \hl{For ECDSA P-256, the 64-byte RRSIG signature together with the fixed RRSIG fields and representative DNS resource-record encoding overhead contribute approximately 104 bytes   $(64 + 40 = 104)$}. Although DNSSEC deployments are increasingly moving toward ECDSA P-256 due to its smaller signatures and operational advantages~\cite{rfc8624,verisign-dnssec-algorithm-update}, we use RSA-2048 as a conservative upper bound. These signatures contribute 256 bytes plus other fields contribute approximately 40 bytes, or
$(256 + 40) = 296$ bytes.
Therefore, \hl{including DNSSEC makes the response size}: $609 + 296 = \mathbf{905}$\textbf{ bytes}.

To support the transaction security of AI agent discovery, agents must bind service endpoints to cryptographic identities. We model this using certificate association data, as in DANE-like mechanisms~\cite{rfc6698}. In such mechanisms, a compact representation of the certificate (typically a cryptographic hash) is stored in DNS along with a small set of control fields that define how the certificate should be interpreted. In practice, SHA-256 is the most widely used matching type, as it is mandatory to implement, while SHA-512 is optional~\cite{rfc7671}. Accordingly, the certificate is represented as a 32-byte SHA-256 hash (or a 64-byte for SHA-512)
\hl{, and with the roughly $35$ bytes associated with control fields, the DNS response size becomes: $905 + 35 = \mathbf{940}$\textbf{ bytes}.}

\begin{figure}[t]
\centering
\includegraphics[width=\linewidth]{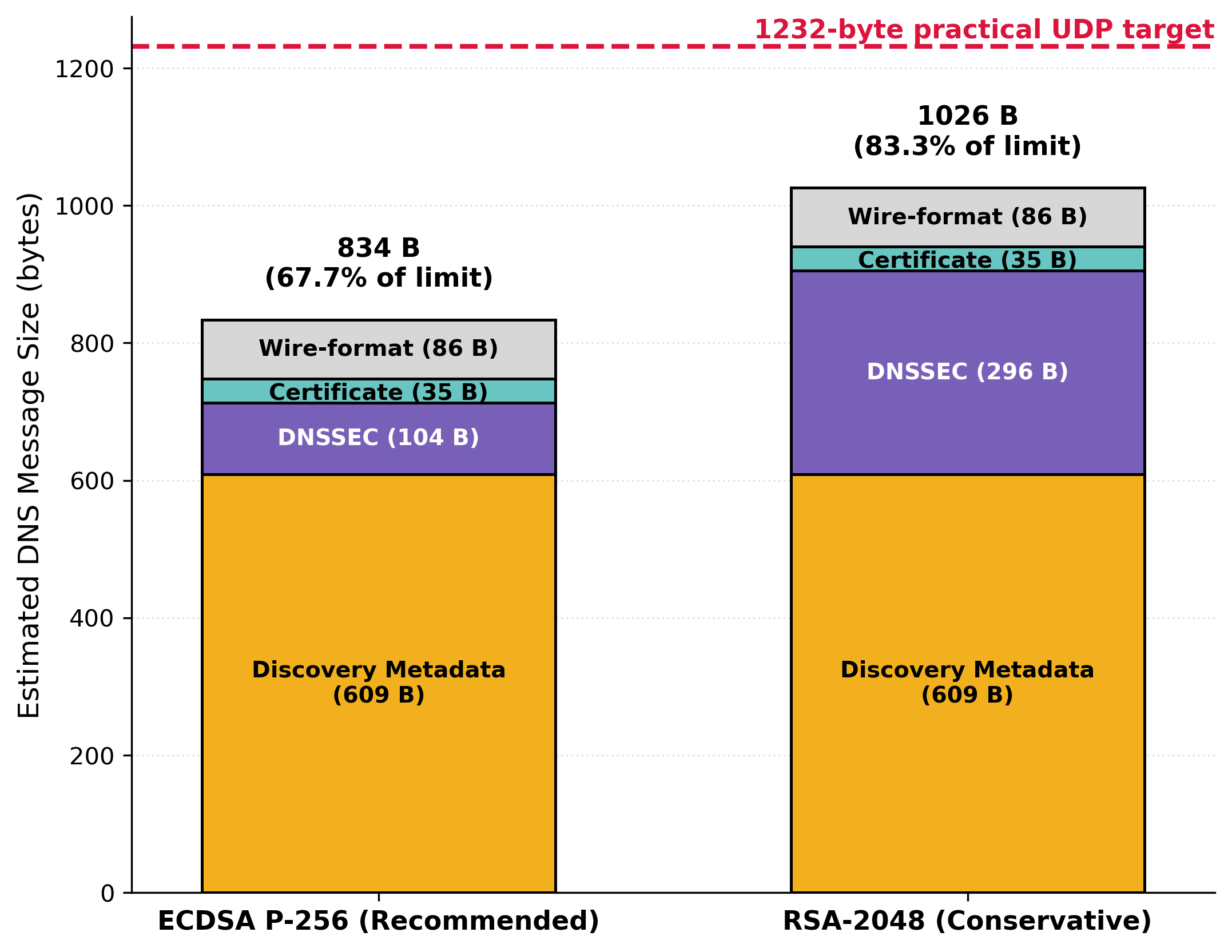}
\caption{\hl{Estimated DNS message size under ECDSA P-256 and RSA-2048 DNSSEC assumptions relative to the practical 1232-byte UDP target.}}
\label{fig:dns-size}
\end{figure}

\hl{Finally, we account for DNS wire-format overhead, including the message header, pre-compression DNS names, and fixed protocol fields, contributing approximately 86 bytes. As summarized in} Table~\ref{tab:size-breakdown}, \hl{the resulting estimated DNS message size is \textbf{1026 bytes}. Because DNS name compression is intentionally excluded, this represents a conservative upper-bound estimate.}
%
\hl{The DNS Flag Day 2020 operational guidance recommends a practical UDP payload target of} 1,232 bytes to avoid IP fragmentation~\cite{dnsflagday-google,rfc8900,goertzen2022}. \hl{Exceeding this target may result in fragmentation or TCP fallback, both of which introduce reliability challenges, with observed} failure rates of approximately 0.1--0.3\% in operational deployments~\cite{dnsflagday-google}.
Figure~\ref{fig:dns-size} illustrates  \hl{estimates of the DNS message size of AI agent discovery under both recommended (ECDSA P-256) and conservative (RSA-2048) DNSSEC assumptions. 
It depicts the discovery overhead
} 
relative to the 1,232-byte UDP response-size target associated with PMTU constraints~\cite{rfc9715,osterweil2009deploying}. 
\hl{These conservative assumptions (summarized in }Table~\ref{tab:size-breakdown}, \hl{)  estimate DNS message sizes of approximately \textit{83.3\%} of the 1{,}232-byte UDP target message-size}.

Our evaluation shows that \hl{the necessary and sufficient discovery} metadata can be \hl{encoded within} \textbf{\textit{\hl{a single DNS resource record}}}, \hl{while remaining within the practical size constraints of a single unfragmented DNS UDP message}. When discovery metadata spans multiple records, tradeoffs arise in query count and resolution latency. Modern DNS deployments support responses of this size through EDNS(0)~\cite{rfc6891}. DNS also provides standardized TCP fallback~\cite{rfc7766} for larger responses when necessary, indicating that metadata size alone does not prevent DNS-based agent discovery.
%
\hl{Furthermore, metadata freshness can be managed by publisher-selected TTL values and DNS operational practices rather than by limitations of the discovery mechanism.}

\textbf{Limitations:}
\hl{This work focuses on AI agents discoverable as Internet services via globally addressable names, consistent with IETF agent discovery approaches. Other interpretations of AI agents may require different discovery models and lifecycle assumptions and are out of scope of this evaluation.} 

This evaluation relies on indirect data sources rather than deployed agentic AI systems. \hl{Capability measurements are derived from static analysis of published MCP server implementations and may not capture dynamically generated capabilities or other runtime behavior. Existing agent tooling ecosystems provide realistic discovery metadata but cannot fully characterize future AI agent deployments. Accordingly, we evaluate metadata size as an initial feasibility condition.}

\textbf{Security Considerations:}
\hl{
Transport security protections can be added if the metadata of an AI agent's capabilities is not intended to be public.
Separately, in some cases DNSSEC's larger messages sizes could warrant consideration of amplification-attack exposure.
Lastly,  DNS-based discovery needs metadata to be consistent between DNS and the agent.
}

\section{Discussion, Future Work, and a Call to Action}

AI agent ecosystems are evolving toward open, cross-platform interaction, where agents must discover and interpret each other without centralized coordination. This creates the need for mechanisms that expose machine-consumable metadata for locating, understanding, and verifying agents.
We argue that an evaluation framework is critical for AI agent discovery\hl{, and} propose three evaluation dimensions: \emph{navigational completeness}, \emph{lookup complexity}, and \emph{transaction performance}. \hl{We} show that the \hl{necessary and sufficient discovery} metadata can \hl{be encoded within a single DNS resource record, fitting within a single DNS UDP message and enabling single-RTT discovery.}

%
As {\bf\emph{future work}}, we will extend this evaluation to a broader set of open-world AI agent discovery systems, including decentralized index-based systems (e.g., Project NANDA~\cite{nanda-ref}), directory and registry-based systems (e.g., AGNTCY~\cite{muscariello2025agntcyagentdirectoryservice}), and naming-based approaches (e.g., DNS-ID~\cite{ihsanullah-dnsid-00}, \hl{DAN}~\cite{seethiraju-dawn-dan-00}, .agent~\cite{dotagent}, AgentDNS~\cite{cui2025agentdnsrootdomainnaming}). \hl{Our novel framework will enable quantitative evaluations of diverse discovery proposals.}

%

We believe now is the time for a {\bf \emph{call to action}}: the new and rapidly evolving AI solution-space needs to be bolstered by the innate (and potentially fallow) power of tried and true modern Internet infrastructure.
We believe that the key to determining which components and aspects of the Internet's infrastructure befit the AI agent discovery design space will be basic evaluation frameworks, such as the one proposed in this paper.
Our call to action is to \emph{not} consider the requirements of AI agent discovery (and its infrastructures, more generally) as requirements for new systems, but rather as integration requirements that may be met or addressed (in whole or in part) by existing research, systems, and infrastructures.
As advances in basic AI and machine learning push the frontiers of research and operations, today's infrastructure is a valuable enabling teacher and resource.
In this regard, the DNS is a venerable protocol that has shown its extensibility throughout its 40 years of life.
It is well suited for making agent metadata discoverable, interoperable, and navigable~\cite{osterweil2012behavior}.
Further, as the inter-domain AI agent discovery space is seeking scalable, resilient, and secure foundations, the DNS' Security Extensions (DNSSEC) represent a 
``terminarch'' (i.e., a 
one-of-a-kind) 
inter-domain security substrate to build on~\cite{osterweil2020cybersecurity}.
Efforts from the research community and basic evaluations will be central to understanding the needs and constructing successful foundations of tomorrow's AI agent architectures.

\bibliographystyle{plain}
\bibliography{references}

@misc{rfc7671,
    series =    {Request for Comments},
    number =    7671,
    howpublished =  {RFC 7671},
    publisher = {RFC Editor},
    doi =       {10.17487/RFC7671},
    url =       {https://www.rfc-editor.org/info/rfc7671},
    author =    {Viktor Dukhovni and Wes Hardaker},
    title =     {{The DNS-Based Authentication of Named Entities (DANE) Protocol: Updates and Operational Guidance}},
    pagetotal = 33,
    year =      2015,
    month =     oct,
}

@misc{rfc6605,
    series =    {Request for Comments},
    number =    6605,
    howpublished =  {RFC 6605},
    publisher = {RFC Editor},
    doi =       {10.17487/RFC6605},
    url =       {https://www.rfc-editor.org/info/rfc6605},
    author =    {Paul E. Hoffman and Wouter Wijngaards},
    title =     {{Elliptic Curve Digital Signature Algorithm (DSA) for DNSSEC}},
    pagetotal = 8,
    year =      2012,
    month =     apr,
}

@misc{rfc8900,
    series =    {Request for Comments},
    number =    8900,
    howpublished =  {RFC 8900},
    publisher = {RFC Editor},
    doi =       {10.17487/RFC8900},
    url =       {https://www.rfc-editor.org/info/rfc8900},
    author =    {Ron Bonica and Fred Baker and Geoff Huston and Bob Hinden and Ole Trøan and Fernando Gont},
    title =     {{IP Fragmentation Considered Fragile}},
    pagetotal = 23,
    year =      2020,
    month =     sep,
}

@misc{rfc9715,
    series =    {Request for Comments},
    number =    9715,
    howpublished =  {RFC 9715},
    publisher = {RFC Editor},
    doi =       {10.17487/RFC9715},
    url =       {https://www.rfc-editor.org/info/rfc9715},
    author =    {Kazunori Fujiwara and Paul A. Vixie},
    title =     {{IP Fragmentation Avoidance in DNS over UDP}},
    pagetotal = 13,
    year =      2025,
    month =     jan,
}

@misc{rfc6891,
    series =    {Request for Comments},
    number =    6891,
    howpublished =  {RFC 6891},
    publisher = {RFC Editor},
    doi =       {10.17487/RFC6891},
    url =       {https://www.rfc-editor.org/info/rfc6891},
    author =    {Joao Luis Silva Damas and Michael Graff and Paul A. Vixie},
    title =     {{Extension Mechanisms for DNS (EDNS(0))}},
    pagetotal = 16,
    year =      2013,
    month =     apr,
}

@misc{rfc7766,
    series =    {Request for Comments},
    number =    7766,
    howpublished =  {RFC 7766},
    publisher = {RFC Editor},
    doi =       {10.17487/RFC7766},
    url =       {https://www.rfc-editor.org/info/rfc7766},
    author =    {John Dickinson and Sara Dickinson and Ray Bellis and Allison Mankin and Duane Wessels},
    title =     {{DNS Transport over TCP - Implementation Requirements}},
    pagetotal = 19,
    year =      2016,
    month =     mar,
}

@misc{rfc9460,
    series =    {Request for Comments},
    number =    9460,
    howpublished =  {RFC 9460},
    publisher = {RFC Editor},
    doi =       {10.17487/RFC9460},
    url =       {https://www.rfc-editor.org/info/rfc9460},
    author =    {Benjamin M. Schwartz and Mike Bishop and Erik Nygren},
    title =     {{Service Binding and Parameter Specification via the DNS (SVCB and HTTPS Resource Records)}},
    pagetotal = 47,
    year =      2023,
    month =     nov,
}

@misc{rfc6698,
    series =    {Request for Comments},
    number =    6698,
    howpublished =  {RFC 6698},
    publisher = {RFC Editor},
    doi =       {10.17487/RFC6698},
    url =       {https://www.rfc-editor.org/info/rfc6698},
    author =    {Paul E. Hoffman and Jakob Schlyter},
    title =     {{The DNS-Based Authentication of Named Entities (DANE) Transport Layer Security (TLS) Protocol: TLSA}},
    pagetotal = 37,
    year =      2012,
    month =     aug,
}

@misc{apisguru-directory,
  title        = {{OpenAPI Directory}},
  author       = {{APIs.guru}},
  url          = {https://github.com/APIs-guru/openapi-directory},
  note         = {https://github.com/APIs-guru/openapi-directory : Accessed: May 2026},
}

@misc{apisguru-index,
  title        = {{API Index}},
  author       = {{APIs.guru}},
  url          = {https://api.apis.guru/v2/list.json},
  note         = {https://api.apis.guru/v2/list.json : Accessed: May 2026},
}

@inproceedings{mockapetris1988development,
  title={{Development of the Domain Name System}},
  author={Mockapetris, Paul and Dunlap, Kevin J.},
  booktitle={Symposium proceedings on Communications architectures and protocols (SIGCOMM)},
  pages={123--133},
  year={1988}
}

@techreport{mozleywilliams-dnsop-dnsaid-01,
    number =    {draft-mozleywilliams-dnsop-dnsaid-01},
    type =      {Internet-Draft},
    institution =   {Internet Engineering Task Force},
    publisher = {Internet Engineering Task Force},
    url =       {https://datatracker.ietf.org/doc/draft-mozleywilliams-dnsop-dnsaid/01/},
    author =    {Jim Mozley and Nic Williams and Behcet Sarikaya and Roland Schott},
    title =     {{DNS for AI Discovery}},
    pagetotal = 25,
    year =      2026,
    month =     mar,
    day =       2,
    note =       {https://datatracker.ietf.org/doc/draft-mozleywilliams-dnsop-dnsaid/01/},
}

@techreport{narajala-courtney-ansv2-01,
    number =    {draft-narajala-courtney-ansv2-01},
    type =      {Internet-Draft},
    institution =   {Internet Engineering Task Force},
    publisher = {Internet Engineering Task Force},
    url =       {https://datatracker.ietf.org/doc/draft-narajala-courtney-ansv2/01/},
    author =    {Scott Courtney and Vineeth Sai Narajala and Ken Huang and Idan Habler and Akram Sheriff},
    title =     {{Agent Name Service v2 (ANS): A Domain-Anchored Trust Layer for Autonomous AI Agent Identity}},
    pagetotal = 44,
    year =      2026,
    month =     apr,
    day =       13,
    note =       {https://datatracker.ietf.org/doc/draft-narajala-courtney-ansv2/01/ : Accessed: May 2026},
}

@misc{rfc4033,
    series =    {Request for Comments},
    number =    4033,
    howpublished =  {RFC 4033},
    publisher = {RFC Editor},
    doi =       {10.17487/RFC4033},
    url =       {https://www.rfc-editor.org/info/rfc4033},
    author =    {Scott Rose and Matt Larson and Dan Massey and Rob Austein and Roy Arends},
    title =     {{DNS Security Introduction and Requirements}},
    pagetotal = 21,
    year =      2005,
    month =     mar,
}

@misc{rfc4034,
    series =    {Request for Comments},
    number =    4034,
    howpublished =  {RFC 4034},
    publisher = {RFC Editor},
    doi =       {10.17487/RFC4034},
    url =       {https://www.rfc-editor.org/info/rfc4034},
    author =    {Scott Rose and Matt Larson and Dan Massey and Rob Austein and Roy Arends},
    title =     {{Resource Records for the DNS Security Extensions}},
    pagetotal = 29,
    year =      2005,
    month =     mar,
}

@misc{rfc4035,
    series =    {Request for Comments},
    number =    4035,
    howpublished =  {RFC 4035},
    publisher = {RFC Editor},
    doi =       {10.17487/RFC4035},
    url =       {https://www.rfc-editor.org/info/rfc4035},
    author =    {Scott Rose and Matt Larson and Dan Massey and Rob Austein and Roy Arends},
    title =     {{Protocol Modifications for the DNS Security Extensions}},
    pagetotal = 53,
    year =      2005,
    month =     mar,
}

@misc{deleg-wg,
    series = {IETF Working Group},
    howpublished = {IETF Datatracker},
    publisher = {IETF},
    author = {{IETF DELEG Working Group}},
    title = {{DNS Delegation (deleg) Working Group}},
    url = {https://datatracker.ietf.org/group/deleg/about/},
    note = {https://datatracker.ietf.org/group/deleg/about/},
}

@misc{dnsflagday-google,
  author       = {{Google Public DNS}},
  title        = {Google Public Experiment Results},
  year         = {2020},
  url          = {https://github.com/dns-violations/dnsflagday/issues/139\#issuecomment-673489183},
  note         = {https://github.com/dns-violations/dnsflagday/issues/139\#issuecomment-673489183 : Accessed: May 2026},
}

@misc{goertzen2022,
  author       = {Jason Goertzen and Douglas Stebila},
  title        = {{Post-Quantum Signatures in DNSSEC via Request-Based Fragmentation}},
  year         = {2022},
  eprint       = {2211.14196},
  archivePrefix= {arXiv},
  primaryClass = {cs.CR},
  url          = {https://arxiv.org/abs/2211.14196},
  note          = {https://arxiv.org/abs/2211.14196},
}

@inproceedings{osterweil2009deploying,
  title={{Deploying and Monitoring DNS Security (DNSSEC)}},
  author={Osterweil, Eric and Massey, Dan and Zhang, Lixia},
  booktitle={{2009 Annual Computer Security Applications Conference (ACSAC)}},
  pages={429--438},
  year={2009},
  organization={IEEE},
}

@misc{nanda-ref,
  author       = {Raskar, Ramesh and Chari, Pradyumna and Zinky, John and others},
  title        = {{Beyond DNS: Unlocking the Internet of AI Agents via the NANDA Index and Verified AgentFacts}},
  year         = {2025},
  eprint       = {2507.14263},
  archivePrefix= {arXiv},
  primaryClass = {cs.NI},
  url          = {https://arxiv.org/abs/2507.14263},
  note = {https://arxiv.org/abs/2507.14263 : Accessed: May 2026},
}

@misc{rfc8624,
    series =    {Request for Comments},
    number =    8624,
    howpublished =  {RFC 8624},
    publisher = {RFC Editor},
    doi =       {10.17487/RFC8624},
    url =       {https://www.rfc-editor.org/info/rfc8624},
    author =    {Paul Wouters and Ondřej Surý},
    title =     {{Algorithm Implementation Requirements and Usage Guidance for DNSSEC}},
    pagetotal = 11,
    year =      2019,
    month =     jun,
}

@misc{verisign-dnssec-algorithm-update,
  author = {{Verisign}},
  title = {{Strengthening Security with DNSSEC Algorithm Update}},
  year = {2023},
  url = {https://blog.verisign.com/security/dnssec-algorithm-update/},
  note = {https://blog.verisign.com/security/dnssec-algorithm-update/ : Accessed: May 2026},
}

@misc{dawn-list,
    series = {IETF Mailing List},
    howpublished = {IETF Mail Archive},
    publisher = {IETF},
    author = {{IETF DAWN Mailing List}},
    title = {{Discussion of Discovery of Agents, Workloads, and Named entities (DAWN)}},
    url = {https://mailarchive.ietf.org/arch/browse/dawn/},
    note = {https://mailarchive.ietf.org/arch/browse/dawn/},
}

@article{osterweil2020cybersecurity,
  title={{A Cybersecurity Terminarch: Use It Before We Lose It}},
  author={Osterweil, Eric},
  journal={{IEEE Security \& Privacy}},
  volume={18},
  number={4},
  pages={67--70},
  year={2020},
  publisher={IEEE}
}

@misc{muscariello2025agntcyagentdirectoryservice,
      title={{The AGNTCY Agent Directory Service: Architecture and Implementation}}, 
      author={Luca Muscariello and Vijoy Pandey and Ramiz Polic},
      year={2025},
      eprint={2509.18787},
      archivePrefix={arXiv},
      primaryClass={cs.AI},
      url={https://arxiv.org/abs/2509.18787}, 
      note = {https://arxiv.org/abs/2509.18787 : Accessed: May 2026},
}

@techreport{ihsanullah-dnsid-00,
    number =    {draft-ihsanullah-dnsid-00},
    type =      {Internet-Draft},
    institution =   {Internet Engineering Task Force},
    publisher = {Internet Engineering Task Force},
    note =      {Work in Progress},
    url =       {https://datatracker.ietf.org/doc/draft-ihsanullah-dnsid/00/},
    author =    {Naveed Ihsanullah},
    title =     {{DNS-Anchored Durable Identity for AI Agents (DNSid)}},
    pagetotal = 36,
    year =      2026,
    month =     apr,
    day =       30,
    note =       {https://datatracker.ietf.org/doc/draft-ihsanullah-dnsid/00/ : Accessed: May 2026},
}

@misc{dotagent,
  author = {{Headless Domains}},
  title = {.agent: Headless Domains for Agentic Applications},
  year = {2025},
  url = {https://headlessdomains.com/agent},
  note = {https://headlessdomains.com/agent : Accessed: May 2026},
}

@misc{cui2025agentdnsrootdomainnaming,
      title={{AgentDNS: A Root Domain Naming System for LLM Agents}}, 
      author={Enfang Cui and Yujun Cheng and Rui She and Dan Liu and Zhiyuan Liang and Minxin Guo and Tianzheng Li and Qian Wei and Wenjuan Xing and Zhijie Zhong},
      year={2025},
      eprint={2505.22368},
      archivePrefix={arXiv},
      primaryClass={cs.AI},
      url={https://arxiv.org/abs/2505.22368}, 
      note = {https://arxiv.org/abs/2505.22368 : Accessed: May 2026},
}

@misc{wu2025mcpzoolargescaledatasetrunnable,
      title={{MCPZoo: A Large-Scale Dataset of Runnable Model Context Protocol Servers for AI Agent}}, 
      author={Mengying Wu and Pei Chen and Geng Hong and Baichao An and Jinsong Chen and Binwang Wan and Xudong Pan and Jiarun Dai and Min Yang},
      year={2025},
      eprint={2512.15144},
      archivePrefix={arXiv},
      primaryClass={cs.CR},
      url={https://arxiv.org/abs/2512.15144}, 
      note = {https://arxiv.org/abs/2512.15144 : Accessed: May 2026},
}

@article{osterweil2012reducing,
  title={Reducing the X. 509 Attack Surface with DNSSEC’s DANE},
  author={Osterweil, Eric and Kaliski, Burt and Larson, Matt and McPherson, Danny},
  journal={Securing and Trusting Internet Names, SATIN},
  volume={12},
  year={2012}
}

@inproceedings{osterweil2012behavior,
  title={Behavior of DNS’Top Talkers, a. com/. net View},
  author={Osterweil, Eric and McPherson, Danny and DiBenedetto, Steve and Papadopoulos, Christos and Massey, Dan},
  booktitle={International Conference on Passive and Active Network Measurement},
  pages={211--220},
  year={2012},
  organization={Springer}
}

@techreport{seethiraju-dawn-dan-00,
    number =    {draft-seethiraju-dawn-dan-00},
    type =      {Internet-Draft},
    institution =   {Internet Engineering Task Force},
    publisher = {Internet Engineering Task Force},
    url =       {https://datatracker.ietf.org/doc/draft-seethiraju-dawn-dan/00/},
    author =    {Ramachandra Rao Seethiraju and Sameer Thakar and Karthik Shyamsunder and Eric Osterweil},
    title =     {{DNS-Based Agent Naming (DAN): AIDISCA and AIINDEX Resource Records for AI Agent Discovery}},
    pagetotal = 21,
    year =      2026,
    month =     jun,
    day =       5,
    note =       {https://datatracker.ietf.org/doc/draft-seethiraju-dawn-dan/ : Accessed: Jun 2026},
}

\end{document}